\documentstyle[prl,aps,floats,twocolumn,graphics]{revtex}

\author{J. Houdayer and O. C. Martin}
\title{Droplet Phenomenology and Mean Field in a Frustrated Disordered System}
\address{Division de Physique Th\'eorique, Institut de Physique Nucl\'eaire, 
Universit\'e Paris-Sud, F--91406 Orsay Cedex, France.}
\date{\today}
\draft

\begin{document}

\maketitle
\begin{abstract}
The low lying excited states of the three-dimensional minimum matching
problem are studied numerically. The excitations' energies grow
with their size and confirm the droplet picture. However, some low energy,
infinite size excitations create multiple valleys in the energy landscape.
These states violate the droplet scaling ansatz, and are consistent with
mean field predictions. A similar picture may apply to spin glasses whereby
the droplet picture describes the physics at small length scales, while mean
field describes that at large length scales.
\end{abstract}


A most useful approach in the study of disordered systems is the replica
method. It has been successfully applied~\cite{Parisi80b} to the Sherrington
and Kirkpatrick (SK) model~\cite{SherringtonKirkpatrick75} of spin glasses,
yielding exact results and revealing remarkable properties such as
multiplicity of nearly degenerate ground states, lack of self-averaging, and
ultrametricity. However it is not clear whether these ``mean field''
properties hold for more realistic spin glass models like the one of Edwards
and Anderson~\cite{EdwardsAnderson75} where finite dimensional effects may
be dominant. To tackle systems in finite dimensions, a number of approaches
based on scaling and the renormalization group have been proposed
\cite{BrayMoore84,McMillan86,FisherHuse88}. In these phenomenological
pictures, it is assumed that there is a unique ground state (up to a global
symmetry) and that excitation energies satisfy a scaling ansatz. For our
purposes, the essential ingredient of this ansatz is that a ``droplet'',
defined as the lowest energy excitation of characteristic size $L$
containing a given spin, is assumed to have an energy which scales as
$L^{\theta}$, with $\theta > 0$. Hereafter we refer to such approaches as
the ``droplet picture''.

Although the ``mean field'' and droplet pictures are very different, they
both agree that there are numerous local minima in the energy landscape
separated by significant energy barriers. The corner-stone of disagreement
between the two approaches concerns the energy of excitations whose size is
comparable to that of the whole system. In the mean field picture, there are
such system-size excitations whose excitation energies are finite, {\it
i.e.}, do not grow with the system size. Thus there are many nearly
degenerate ground states and the energy landscape consists of numerous
similar low energy valleys. On the contrary, in the droplet picture, the
characteristic energy for such system-size excitations grows as a positive
power of the size of the system. As a consequence, the probability of having
such an excitation with an energy below a fixed value goes to zero as the
system size grows. Thus the ground state is almost never nearly degenerate
with another significantly different local minimum. Furthermore, from the
point of view of the droplet scaling ansatz, the existence of many nearly
degenerate ground states would lead to $\theta \le 0$, and yet, for the spin
glass phase to exist at non-zero temperatures, droplet excitations must be
suppressed, leading to $\theta > 0$.

These points show that an unambiguous determination of the lowest energies
of large scale excitations would help resolve the controversy over the
relevance of the droplet and mean field pictures to finite dimensional spin
glasses. Unfortunately, the main obstacle in the way of such a test is the
computational complexity of spin glasses: just finding the ground state is
{\it NP}-hard~\cite{PapadimitriouSteiglitz82}, and finding the excited
states is at least as demanding computationally. We have thus chosen to
investigate a different system: the three-dimensional minimum matching
problem (see below). Although it is both frustrated and disordered, it is
computationally more tractable than a spin glass. For this system we have
devised a new algorithm which allows us to enumerate very efficiently {\it
all} the states above the ground state in a systematic way. To our
knowledge, this is the first time it is possible to explore unambiguously a
non-trivial frustrated disordered model. With our computational tool and
some analysis, we find that the droplets in this model have energies which
grow with their size, justifying a droplet picture with a positive exponent
$\theta$. However we also find that the thermodynamic limit is a bit
singular. In particular, some ``infinite'' size droplets appear at low
energies, creating an energy landscape with many nearly degenerate valleys.
Our three-dimensional model thus has a droplet like behavior at finite
length scales, but its energy landscape at large length scales is as
predicted by mean field.

\paragraph*{The model ---}
We consider a system arising in combinatorial optimization: the minimum
matching problem (MMP)~\cite{PapadimitriouSteiglitz82}. This choice 
is motivated by the following properties: 
(i) the problem of finding ground states of 
two-dimensional spin glasses can be mapped to a MMP~\cite{BiecheMaynard80}
(see~\cite{Bendisch97} for recent developments);
(ii) we are able to compute quickly and exactly
the ground state {\it and} the excited states of the MMP for any realization
of the disorder; (iii) a droplet picture can be constructed quite naturally;
(iv) the replica approach has been used to solve a mean field approximation
of the model~\cite{MezardParisi86}.

Consider $N$ points ($N$ even) and the set of ``distances'' between them.
These distances define the instance, that is the quenched disorder. A
micro-state or configuration is any ``matching'' of the points, that is a
dimerization, so that each and every point is paired with exactly one other
point. The energy of a matching is defined as the sum of the distances
between matched points. The MMP, as defined in combinatorial optimization,
consists of finding the lowest energy matching. Because of the disorder, the
system is frustrated: each point would like to be paired to its nearest
neighbor, but in general this cannot be achieved for all points. In the {\it
statistical physics} formulation, one sums over all micro-states, weighting
them with the Boltzmann factor. One also takes the thermodynamic limit
($N\to \infty$), and for that, one must specify the quenched disorder
ensemble.

In the Euclidean form of the MMP, one considers $N$ points at random in a
$3$-dimensional volume. In order to avoid edge effects, we use a cube with
periodic boundary conditions. The large $N$ limit is taken at fixed density
of points, and thus corresponds to the usual infinite volume limit. (In
these units, which we use hereafter, the volume is equal to $N$
and the ground state energy is extensive, {\it i.e.},
proportional to $N$.) To tackle
this model, it is useful to consider a mean field approximation of the
problem; this has been done by M\'ezard and Parisi~\cite{MezardParisi86} who
applied replicas to a modified model where all correlations among distances
between points were removed. Hereafter, we call this modified model the
(independent) {\it random-link} MMP because all the ``distances'' between
pairs of points are independent random variables. These individual distances
are taken to be distributed as in the Euclidean model.

In a parallel with spin glasses, one can consider the Euclidean MMP to be
the analogue of the Edwards-Anderson model, incorporating frustration and
disorder in a Euclidean space; similarly, the analogue of the SK (or better
yet, the Viana-Bray~\cite{VianaBray85}) model is the {\it random-link}
MMP. It is known that the random-link model provides an excellent
approximation for the ground state energy density of
the Euclidean MMP. (See~\cite{HoudayerBoutet98} for an overview of some
of the associated properties of ground states.)
Now we will see that it also enables one to understand
the corresponding energy landscape.

\begin{figure}
\begin{center}
\resizebox{0.7\linewidth}{!}{\includegraphics{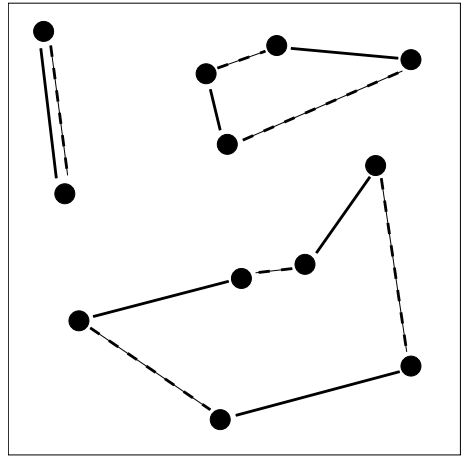}}
\end{center}
\caption{Comparison of two matchings. One is in solid lines,
the other is in dashed lines. Non intersecting alternating loops 
describe the difference of the two matchings. Here 
there are two loops, of sizes 4 and 6.}
\label{FigLoop}
\end{figure}

\paragraph*{Excitations ---}
A general matching differs from the ground state by replacing some of the
bonds by others. One can organize those two sets of bonds into alternating
loops where every other bond in a loop belongs to the ground state matching,
the others belonging to the excited state matching. (This follows from the
fact that each point is matched to one and only one other point; see
Figure~\ref{FigLoop} and~\cite{GrassbergerFreund90}.) Any excited state thus
consists of alternating loops which are non-overlapping,
{\it i.e.}, which have no points in common. 
The total excitation energy is the
sum of the energies of each loop. With this insight, the system can be
viewed as an (interacting) gas of loops. Hereafter, we consider only
excitation energies, {\it i.e.}, all energies are measured relative to the
ground state energy. Also, we define the {\it size} of a loop as the number
of bonds it has; the size $\ell$ of any loop is thus even and satisfies
$4\leq\ell\leq N$.

\paragraph*{Local density of states ---}
At low enough temperatures, this gas of loops becomes very dilute,
suggesting that ``loop-loop'' interactions may be neglected (recall that the
loops cannot overlap). In this approximation, the thermodynamics of the
gas may be computed from the local density of states associated with single
loop excitations. Thus we have determined numerically the local density
$\rho_{\ell}(E)$ of one-loop states of size $\ell$ and of energy $E$. 
To accomplish this, for each realization of the disorder, we
generate all the single loops up to a maximum energy. 
Our algorithm does this in a systematic way 
by successively increasing the length of
different bonds and finding the new ground states. This increase in length
has the effect of preventing the new ground states from containing
certain bonds. The process can be organized into a tree search with
a branch and bound so that all the states below a given energy
are obtained. We then determine the
number ${\cal N}_{\ell}(E)$ of loops of size $\ell$ with energy between $E$
and $E + \Delta E$. We have averaged ${\cal N}_{\ell}(E)$ over $10^3$ to
$10^4$ randomly generated instances for $N = 50, 100, 150, 200$, and $250$
points; this leads to the estimator $\rho_{\ell}(E)=\langle {\cal N}_{\ell}
\rangle / N \Delta E$.
Our data show that the different values of $N$ lead to the same function,
justifying the definition of $\rho_{\ell}(E)$ and indicating that our values
of $N$ are large enough for finite size effects to be negligible. For each
$\ell$, we find that $\rho_{\ell}(E)$ is a smoothly increasing function of
$E$ with the property $\rho_{\ell}(0) \ne 0$. This fact can be understood by
considering the measure of the points leading to a loop of zero energy. (It
should be clear that finite dimensional spin glasses also have this property
because the probability density of having a cluster of spins in a null local
field is non-zero.)

\paragraph*{Droplets ---}
Fisher and Huse~\cite{FisherHuse88} define droplets in the context of spin
glasses; generalizing their definition to the matching problem is
straightforward. For a given point, consider the set of all single loops of
size $\ell$ ($L \leq \ell < 2L$) passing through that point. Define the
droplet of characteristic size $L$ containing that point as the loop in the
defined set with the lowest energy. When applied to the MMP, the droplet
picture states that the typical energy of droplets of size $L$ scales as
$L^{\theta}$, $\theta > 0$. Furthermore, the scaling
ansatz~\cite{FisherHuse88} says that the probability distribution of the
energy $E_L$ of a droplet of size $L$ behaves as
\begin{equation}
\label{eq_P_L}
P_L(E_L) = p(E_L/L^{\theta}) / L^{\theta}
\end{equation}
with $p(0) \neq 0$. A direct test of this scaling ansatz is beyond the
possibilities of our numerics because large values of $L$ would require too
large computation times. Thus we have instead performed an indirect test of
the scaling ansatz as follows.

Our method is based on relating $p(0)$ to the $\rho_{\ell}(0)$'s. Since we
are concerned with very low energies, excitations are rarefied; as in the
dilute gas approach, we will assume that the excitations are independent.
Using Equation~\ref{eq_P_L}, we derive the probability distribution of the
lowest energy droplet of size $L$ in the {\it whole} system, and find that
the mean of this distribution is $L^{1+\theta} / N p(0)$. (For this, we
assumed that the droplets were independent; we also used the property that
the number of droplets of size $L$ is $N/L$ up to constant factors, each
droplet containing $O(L)$ points.) We can also calculate this mean using the
$\rho_{\ell}(0)$'s; setting the two expressions to be equal leads to the sum
rule
\begin{equation}
\label{eq_rhos}
\rho_{L}(0) + \rho_{L+2}(0) \cdots + \rho_{2L-2}(0) = p(0) / L^{1 + \theta} .
\end{equation}
(In the case of spin glasses, the exponent would be $3 + \theta$ assuming
that droplets of size $L$ have $O(L^3)$ spins.) If the scaling with $\theta
> 0$ is valid, Equation~\ref{eq_rhos} gives $\rho_{\ell}(0) =
O(\ell^{-2-\theta})$. In Table~\ref{TableDroplet} we give the results for
$\rho_{\ell}(0)$ and $\ell^2 \rho_{\ell}(0)$ as a function of $\ell$ for the
values of $\ell$ within reach of our computations. (Our data have
statistical errors which prevent us from going to much larger values of
$\ell$ in a meaningful way.) The positivity of $\theta$ is confirmed by the
decrease of these quantities with increasing $\ell$, giving good evidence
that the droplet picture applies to the MMP.

\begin{table}
\caption{Density of states at zero energy as a function of the 
loop size in the three dimensional Euclidean MMP.}
\label{TableDroplet}
\begin{tabular}{ddd}
$\ell$ & $\rho_{\ell}(0)$ & ${\ell}^2 \rho_{\ell}(0)$\\
\hline
4 & 4.520 & 72.3\\
6 & 0.840 & 30.2\\
8 & 0.367 & 23.5\\
10 & 0.205 & 20.5\\
\end{tabular}
\end{table}

\begin{figure}
\begin{center}
\resizebox{0.95\linewidth}{!}{\includegraphics{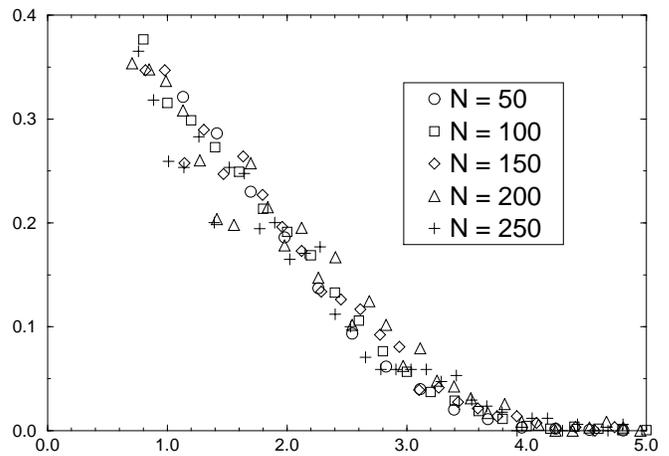}}
\end{center}
\caption{Scaling function $G(x)$ (see Equation~\ref{EquScaling}) for different
values of $N$.}
\label{FigScaling}
\end{figure}

\paragraph*{Breakdown of the droplet picture ---}
Consider now the distribution $P(\ell{_1},N)$ of the length $\ell{_1}$ of
the first excited state. This quantity is easy to extract numerically;
furthermore, $P(\ell{_1},N)$ is a probability distribution over $\ell{_1}$,
and a calculation similar to the one just discussed gives $P(\ell{_1},N) =
\rho_{\ell_1}(0) / \sum_{\ell '} \rho_{\ell '}(0)$. When $N \to \infty$,
$P(\ell{_1},N)$ converges pointwise to a limiting distribution which falls
off quickly with $\ell{_1}$; the nature of this fall off is consistent with
the droplet picture as we saw previously. However, we also find that
anomalously large loops appear with frequency $O(1/{\sqrt{N}})$. These large
loops have lengths which grow as $\sqrt{N}$, and their distribution
satisfies the following scaling law as $N \to \infty$
\begin{equation}
\label{EquScaling}
Prob(\tilde{\ell}_1 = x) \sim G(x) / \sqrt{N}
\end{equation}
with the scaling variable $\tilde{\ell}_1 = \ell_1/\sqrt{N}$. This scaling
is illustrated in Figure~\ref{FigScaling}; the finite $x$ contributions at
different (large) values of $N$ lead to the same curve $G(x)$. (Note that
the fixed size loops lead to a delta function contribution at $x=0$.)
This scaling is incompatible with the droplet picture as can be seen by
considering the moments of $P(\ell{_1},N)$. If the exponent $\theta$
existed, $\rho_{\ell}(0)$ (and thus $P(\ell{_1},N)$) would be
$O(\ell^{-2-\theta})$. Then the moments $\langle {\ell_1}^{1 + \delta}
\rangle$ would be finite for $\delta < \theta$ and would diverge as $N \to
\infty$ for $\delta > \theta$.
However, from Equation~\ref{EquScaling}, the divergence sets in as soon as
$\delta > 0$ because of the contribution from the anomalously large loops.
The conclusion is that although the droplet picture shows all signs of being
correct when one takes the limit $N \to \infty$ while keeping the scale
fixed, it is not valid if one considers scales which grow with the system
size!

\paragraph*{Mean field picture and energy landscapes ---}
To shed light on these anomalously large loops, consider the mean field
picture as obtained by using the properties of the random link MMP. For that
model, we have repeated the calculations performed in the Euclidean case and
have determined spectra of energies and the sizes of the corresponding
excitations. (Although the random link model has been solved by the replica
method, this kind of information has not been obtained previously.) First of
all, we find that all low lying excitations have sizes of $O(\sqrt{N})$. In
particular, for the first excited state, we find that the large $N$ scaling
is given by $Prob(\tilde{\ell}_1 = x) \sim G_{RL}(x)$, again with
$\tilde{\ell}_1 = \ell_1 / \sqrt{N}$. The random link model excitations thus
have the same $\sqrt{N}$ scaling in size as the anomalous excitations in the
Euclidean model, and in fact, the scaling functions $G(x)$ and $G_{RL}(x)$
are qualitatively similar. Note that $G_{RL}(x)$ has no delta function peak
at $0$, {\it i.e.}, no contribution from finite size loops; this can be
understood from the ``geometry'' of the random link model: its structure is
locally that of a Cayley tree, and as $N \to \infty$, finite size loops
connecting near neighbors disappear. Second of all, we find the following
scaling law
\begin{equation}
\label{eq_rl_scaling}
\langle{\cal N}_{RL}(\tilde{E})\rangle/\Delta\tilde{E}\sim R_{RL}(\tilde{E}) 
\end{equation}
where $\tilde{E}=E\sqrt{N}$ and ${\cal N}_{RL}(\tilde{E})$ is the number of
loops of (rescaled) energy between $\tilde{E}$ and
$\tilde{E}+\Delta\tilde{E}$; the scaling function $R_{RL}$ increases like an
exponential.

These results show that the random link MMP has many low-energy large-scale
excitations, the characteristic size of which is $O(\sqrt{N})$ and the
characteristic energy of which is $O(1/\sqrt{N})$. To obtain a mean field
picture for the Euclidean model, we can say that the large scale excitations
of the random link model ``survive'' in the Euclidean model; if we add the
small size droplets to these large scale excitations, we generate valleys.
(Note that essentially the same droplets make up the different valleys, so
that the valleys are nearly identical in structure.) If this picture is
correct, we expect the statistics of the excitations associated with the
bottom of the valleys of the Euclidean model to be qualitatively similar to
the statistics of the states in the random link model. To better see these
``valley'' states, we have studied the loops of size greater than
$C\sqrt{N}$ (where $C$ is a constant). With this restriction, we find that
the scaling in size and energy of low energy excitations is very similar in
the random link and Euclidean models. In particular, the Euclidean model
satisfies Equation~\ref{eq_rl_scaling} with a scaling function $R$ which is
close to $R_{RL}$.

In view of the fact that there is no replica symmetry breaking in the MMP,
it may seem surprising to have such a structured landscape; one would expect
instead the droplet picture to be the whole story. But this is not the case,
the droplet picture breaks down on the scales where the mean field picture
predicts large scale excitations of low energy. Since these excitations
involve only $O(\sqrt{N})$ bonds, the valleys have an overlap which tends
towards one in the $N \to \infty$ limit, and this is consistent with the
absence of replica symmetry breaking. But the point is that these valleys
differ by an infinite number of bonds in that limit, and that the droplet
picture is valid only within a single valley.

Let us speculate on how our results may extrapolate to the case of the
Edwards-Anderson spin glass model. There may be two types of low energy
excitations: the first given by the droplet picture and associated with
fixed sizes, and the second given by the mean field picture and associated
with system-size excitations involving $O(N)$ spins. This second
contribution is responsible for the valleys in the energy landscape. The
bottom of the valleys can be thought of as states similar to those arising
in the mean field picture, having statistics well described by that
approach. In particular, the statistics of the bottom of those valleys may
well obey a scaling law such as Equation~\ref{eq_rl_scaling} with ${\tilde
E} = E N^\gamma$ where $\gamma$ is a new exponent; the characteristic
inter-level spacing of these valley energies is then $O(N^{- \gamma})$. Of
course, the small size droplets give rise to energy levels with a
characteristic inter-level spacing of $O(1/N)$. We expect the exponent
$\gamma$ to be given exactly by mean field. 
We are currently investigating this question.

\acknowledgments
We are grateful to O. Bohigas, D. Dean and H. Hilhorst for their detailed
comments. J.H. acknowledges a fellowship from the MENESR, and O.C.M.
acknowledges support from the Institut Universitaire de France. The Division
de Physique Th\'eorique is an Unit\'e de Recherche des Universit\'es
Paris~XI et Paris~VI associ\'ee au CNRS.

\bibliographystyle{prsty}
\bibliography{/tmp_mnt/home/houdayer/Papers/Biblio/references}

\end{document}